\documentstyle[aps,prl,twocolumn,epsf]{revtex}

\begin{document}
\draft



\twocolumn[\hsize\textwidth\columnwidth\hsize\csname %
@twocolumnfalse\endcsname
%


\newcommand{\lsix}{La$_{1.94}$Sr$_{0.06}$CuO$_4$}
\newcommand{\lsco}{La$_{2-x}$Sr$_x$CuO$_4$}
\newcommand{\la}{$^{139}$La}
\newcommand{\cu}{$^{63}$Cu}
\newcommand{\cuo}{CuO$_2$}


\title{Charge Segregation, Cluster Spin-Glass and Superconductivity in \lsix}


\author{M.-H. Julien$^{1,2}$, F. Borsa$^{1,2}$, P. Carretta$^1$,
M. Horvati\'c$^{3}$, C. Berthier$^{3,4}$ and C.T. Lin$^{5}$}


\address{$^1$Dipartimento di Fisica "A. Volta", Unit\'a INFM di Pavia,
Via Bassi 6, I-27100 Pavia, Italy}
\address{$^2$Department of Physics, Ames Laboratory, Iowa State University, Ames IA-50011}
\address{$^3$Grenoble High Magnetic Field Laboratory, CNRS and MPI-FKF,
BP 166, F-38042 Grenoble Cedex 9, France}
\address{$^4$Laboratoire de Spectrom\'etrie Physique, Universit\'e
J. Fourier, BP 87, F-38402 St. Martin d'H\`eres, France}
\address{$^5$Max-Planck-Institut f\"ur Festk\"orperforschung, Heisenbergstrasse 1,
D-70569, Stuttgart, Germany}



\date{February 10, 1999}
\maketitle


\widetext



\begin{abstract}


A \cu~and \la~NMR/NQR study of superconducting ($T_c$=7~K) \lsix~single crystal is reported.
Coexistence of spin-glass and superconducting phases is found below $\sim$5~K
from \la~NMR relaxation.
\cu~and \la~NMR spectra show that,
upon cooling, CuO$_2$ planes progressively separate into two 
magnetic phases, one of them having enhanced antiferromagnetic 
correlations. 
These results establish the AF-cluster nature of the spin-glass.
We discuss how this phase can be related to the
microsegregation of mobile holes and to the possible pinning of charge-stripes.


\end{abstract}



\pacs{PACS numbers: 76.60.-k, 74.25.Ha, 74.72.Dn}
]
\narrowtext



\newcommand{\lsix} {La$_{1.94}$Sr$_{0.06}$CuO$_4$}
\newcommand{\lsco} {La$_{2-x}$Sr$_x$CuO$_4$}
\newcommand{\ybco} {YBa$_{2}$Cu$_3$O$_{6+x}$}
\newcommand{\la}   {$^{139}$La}
\newcommand{\cu}   {$^{63}$Cu}
\newcommand{\cuo}  {CuO$_2$}
\newcommand{\etal} {{\it et al.}}


Although \lsco~is one of the most studied and structurally simplest
high-$T_c$ superconductor,
the complexity of its phase diagram keeps increasing every year. 
A striking feature is that,
while N\'eel antiferromagnetic (AF) order
is fully destroyed by $x$=2 \% of doped holes, samples with much higher
doping still show clear tendencies towards spin ordering:


- At intermediate concentrations
between N\'eel and superconducting phases (0.02$\leq$$x$$\leq$0.05), 
a spin-glass phase is found \cite{Harshman88,Sternlieb90,Chou95}.
There are indications, but no direct evidence, that this phase is 
formed by frozen AF clusters, which could originate from
the spatial {\it segregation} of doped holes in CuO$_2$ planes: a "cluster spin-glass"
\cite{Hayden91,Cho92,Chou93,Emery93,Kivelson98,Gooding97,Niedermayer98}.
Strikingly, this spin-glass phase is found to coexist with superconductivity
\cite{Niedermayer98} (see also \cite{Kitazawa88,Weidinger89}).


- Commensurability effects around
$x$=0.125 (=1/8) and/or subtle structural modifications help restoring 
long-range AF order. This is also understood as a consequence of segregation of
doped-holes, but
here charges are observed to order into 1D domain walls, or "stripes" \cite{Tranquada95}.
Again magnetic order is claimed to coexist with bulk superconductivity
\cite{Tranquada97,Ostenson97,Suzuki98,Nachumi98}.


Clearly, the context of static magnetism and charge segregation in which superconductivity
takes place is the central question in this region of the phase diagram
\cite{Emery93,Kivelson98,Gooding97,Theory}. So,
a lot should be learnt from the microscopic nature of the cluster spin-glass phase, which
has not been clarified yet, and from the passage from spin-glass to superconducting
behaviour.


Here, we address this problem through a comprehensive nuclear magnetic resonance (NMR)
and nuclear quadrupole resonance (NQR) investigation of \lsix,
a compound at the verge of the (underdoped) superconducting phase ($T_c$=7~K).
In addition to the confirmation of coexisting spin-glass and superconducting phases,
the AF-cluster nature of the spin-glass is microscopically
demonstrated from \cu~and \la~NMR spectra. 
We discuss how the observed microscopic phase separation can be related to the
microsegregation of mobile holes in CuO$_2$ planes, and
suggest that the cluster spin-glass is the magnetic counterpart of a pinned,
disordered, stripe phase: a "stripe-glass" \cite{Kivelson98}. 


The sample is a single crystal ($\sim$200 mg), grown from solution as described in Ref.
\cite{Lin97}.
Magnetization measurements have shown a superconducting transition
with an onset at $T_c$=7 K.


We first discuss the NQR measurements.
The $^{63}$Cu nuclear spin-lattice relaxation rate 1/$^{63}T_1$ was measured
at the center of the NQR line shown in Fig. 1(a).
The recovery of the magnetization after a sequence
of saturating pulses, was a single exponential at all temperatures. The results are shown
in Fig. 1(b) \cite{CommentT1}.
It is remarkable that for the same hole concentration and a similar $T_c$,
we obtain identical Cu NQR spectra (central frequency, width, and small high frequency tail
from the anomalous "B" line -sites with a localized doped-hole
\cite{Fujiyama97,Hammel98})
and the same $^{63}T_1$ values as Fujiyama \etal~\cite{Fujiyama97}.
All these quantities are strongly doping-dependent.
This is a very good indication of the precision and the homogeneity of the Sr
concentration in our sample $x$=0.06$\pm$0.005.
Below 250~K, 1/$^{63}T_1$ flattens and it decreases below $\sim$150~K.
This regime could not, however, be 
explored since the Cu nuclear spin-spin relaxation time ($T_2$) shortens drastically
upon cooling, making the NMR signal too small for reliable measurements,
especially below $\sim$50~K.


A useful substitute of \cu~measurements is the NQR/NMR of \la.
Although La lies outside CuO$_2$ planes, it is coupled to Cu$^{2+}$ spins
through a hyperfine interaction, whose magnitude is small compared
to that on \cu, leading to a long value of $^{139}T_2$.
A typical \la~NQR line (3$\nu_Q$ transition) is shown in Fig. 1(c).
The asymmetry is perfectly accounted for by a two-gaussian fit, which is very
similar to that found in stripe-ordered
La$_{1.48}$Nd$_{0.4}$Sr$_{0.12}$CuO$_4$ \cite{Teitelbaum98}.
The existence of two electric field gradient contributions is 
related to static charge inhomogeneities, either directly and/or indirectly
through different tilt configurations of CuO$_6$ octahedra.


\vspace{-0.7cm}
\begin{figure}[b!]
\begin{center}
 \epsfxsize=95mm
 $$\epsffile{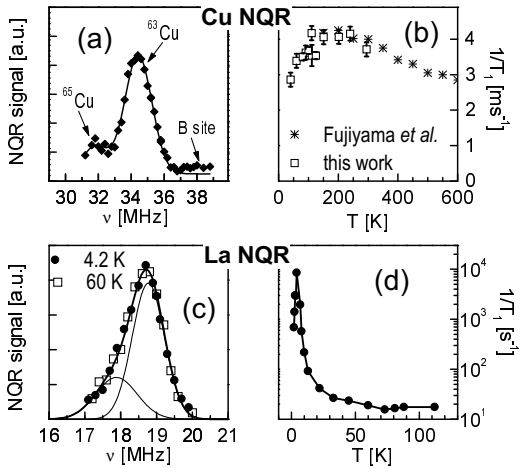}$$
\end{center}
\caption{(a) $^{63,65}$Cu NQR spectrum. (b) NQR $^{63}$Cu 1/$T_1$; this study (squares)
and from ref. \protect\cite{Fujiyama97} (stars). (c) $^{139}$La NQR spectrum, decomposed
into the sum of two gaussians. (d) NQR $^{139}$La 1/$T_1$ showing the spin-freezing transition.}
\label{NQR}
\end{figure}
By comparing the recovery law of the \la~magnetization after saturation of
the 2$\nu_Q$ transition with that measured on the 3$\nu_Q$ transition, it was found
that the spin-lattice relaxation 
is due to both magnetic {\it and} electric field gradient fluctuations around 100~K. However,
below 75~K 1/$T_1$ increases progressively upon cooling and becomes
entirely of magnetic origin. As seen in Fig. 1(d), 1/$T_1$ increases by
almost three orders of magnitude with a peak at $T_g$$\simeq$5~K. This behaviour
is typical of a slowing down of spin-fluctuations, 1/$T_1$ reaching
a maximum when the frequency of these fluctuations is equal to the nuclear
resonance
frequency, here $\nu_Q$$\simeq$18~MHz (or equivalently a correlation time
$\tau$$\sim$10$^{-8}$s).
Thus, a spin-freezing occurs {\it in the superconducting state} of \lsix.
This adds a new item to the list of unconventional properties of the cuprates,
which have to be addressed by any theory.
The scale, microscopic or mesoscopic, on which both types of order
coexist is a crucial question which cannot be addressed here.
But we stress again that our results are representative of 
an homogeneous x=0.06 Sr concentration. This is also confirmed by the value $T_g$$\simeq$5~K,
which is in quantitative agreement with the carefully established NQR \cite{Chou93}
and $\mu$SR phase diagrams \cite{Niedermayer98} of \lsco~
(the characteristic times of NQR and $\mu$SR are similar).
The freezing process is characterized by a high level of
inhomogeneity, since a very wide distribution of $T_1$ values
develops below 50~K, as inferred from the streched exponential time decay of
the nuclear magnetization \cite{T1}.
As already noticed in Refs. \cite{Harshman88,Hayden91}, the slowing down starts around 70~K,
in the temperature range where the in-plane resistivity $\rho_{ab}$ has a minimum.
Thus, charge localization seems to be a precursor effect of Cu$^{2+}$ spin freezing.


It is also important to probe the {\it local} static magnetization in CuO$_2$ planes.
This can be characterized through
the shift $K_{cc}$ (for $H_0$$\|$$c$) of the \cu~NMR line
which is the sum of a {\it $T$-independent} orbital term
$K^{\rm orb}\simeq$1.2\% plus a contribution from the spin susceptibility:
\begin{equation}
^{63}K_{cc}^{\rm spin}=\frac{(A_{cc}+4B)}{g_{cc}\mu_B} \frac{<S_z>}{H_0} .
\label{shift1}
\end{equation}
$A_{cc}$ is the hyperfine coupling with on-site electrons,
$B$ the transferred hyperfine coupling
with electrons on the first Cu neighbour, $g$ the Land\'e factor, 
and $<$$S_z$$>$ the on-site Cu moment,
here assumed to be spatially homogeneous on the scale
of the Cu-Cu distance. Since $A_{cc}+4B\simeq$0 in \lsco~
and \ybco, one usually has negligible magnetic shift $^{63}K_{cc}^{\rm spin}$$\simeq$0.


The inset to Fig. 2 shows the $^{63}$Cu NMR central line at room 
temperature.
There are clearly two contributions: a relatively sharp line with 
the usual shift $K_c$$\sim$1.2\%,
and a slightly shifted, much broader, background. 
The perfect overlap of the NMR intensity {\it vs.} shift plots at 17 and 24 Tesla
asserts that the broadening
is purely magnetic, {\it i.e.} it is a distribution of
shifts $K_c$.
This distribution is considerable ($\pm$2-3\%),
exceeding by far anything ever seen in the cuprates.
Also striking is the $T$-dependence of the spectrum (Fig. 2).
The NMR signal clearly diminishes upon cooling.
The effect is more dramatic for the main peak, which disappears between 100
and 50~K.
At 50~K, the spectrum is only composed of a background, at least two times wider
than at 300~K. 
The shortening of $T_2$ (by a factor of two from 300~K to 100~K)
accounts for a small fraction of the intensity loss.
Some signal is redistributed from the main peak to the background signal,
but part of it is actually not observed, due to the huge spread of resonance
frequencies.


It is evident from Eqn. 1 that $K_c$$\neq$0 values are
possible {\it only} if $<$$S_z$$>$
is strongly spatially modulated on scale of {\it one} lattice spacing,
so that the shift for a Cu site at position $(x,y)$
cannot be written as in Eqn. \ref{shift1}, but contains the sum of terms:
$A_{cc}$$<$$S_z(x,y)$$>$+$B$[$<$$S_z(x\pm 1,y)$$>$+
$<$$S_z(x,y\pm 1)$$>$].
In fact, large values of $K_c$ such as found here imply
that the local magnetization is {\it staggered}:
the cancellation of the $A_{cc}$$<$0 and $B$$>$0 terms in Eqn.~1
is removed by the sign alternation of $<$$S_z$$>$ from one site
to its nearest Cu neighbours, thus allowing $|K_c|$$\gg$0 locally.


The presence of substantial staggered magnetization is striking. One way to generate
such enhanced AF correlations could be that some localized doped-holes act as static defects
in the magnetic lattice, somehow similar to the substitution of Zn for Cu \cite{CommentZn}.
However, only one broadened peak is detected in Cu NMR studies of Zn-doped YBCO,
while there are here two well-defined magnetic phases (see also \la~results below):
Furthermore, there is some staggered magnetization already at 290~K, where $\rho_{ab}$
is metallic-like, and the \cu~NQR B site, which is known to be related to
localized holes \cite{Hammel98}, is extremely small here. So,
an impurity-like effect from localized holes does not explain the data.
\vspace{-0.7cm}
\begin{figure}[b!]
\begin{center}
 \epsfxsize=80mm
 $$\epsffile{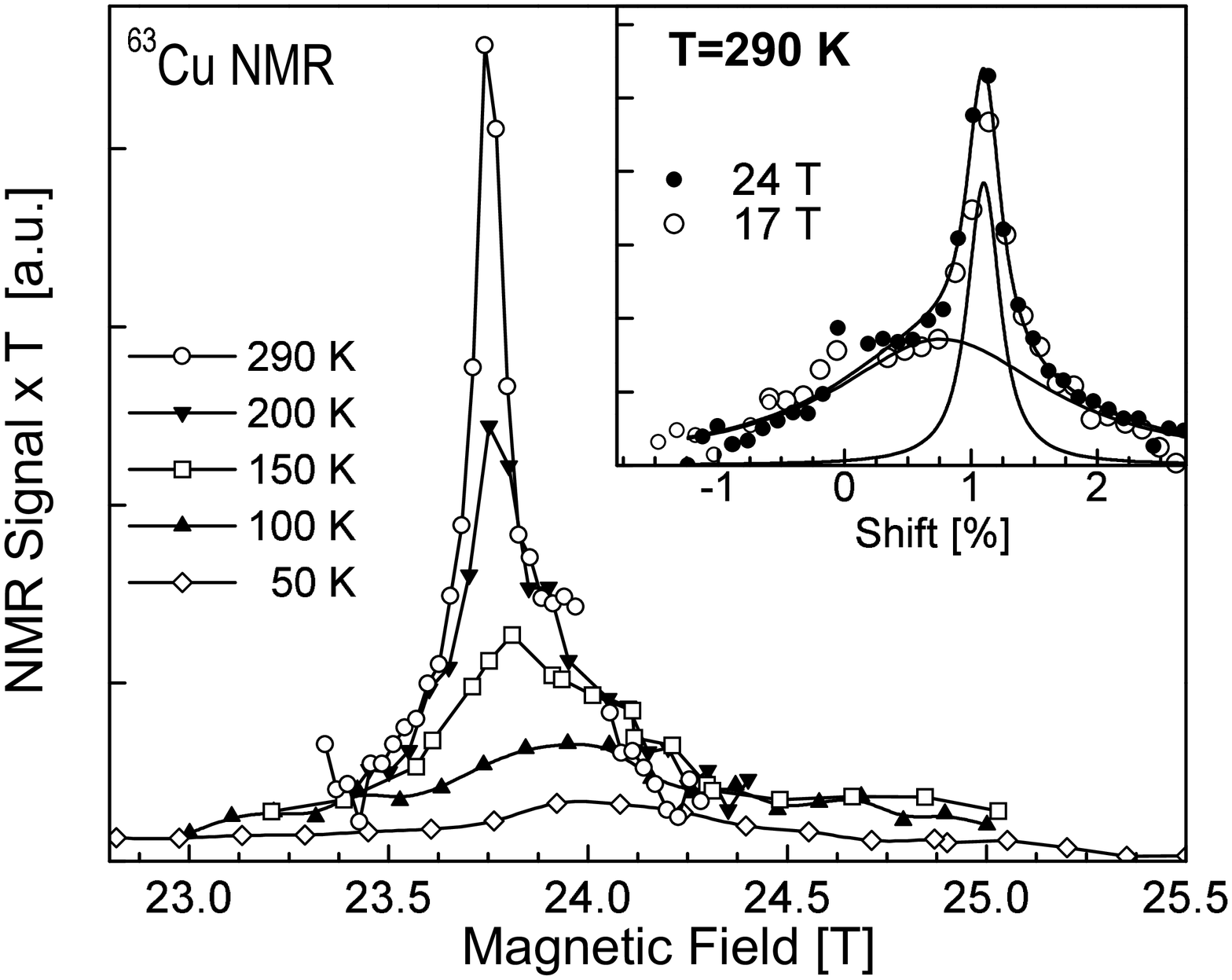}$$
\end{center}
\caption{Main panel: Field swept $^{63}$Cu NMR spectra as a function of $T$
($H$$\|$$c$), recorded in the very same experimental
conditions. The relative intensities can thus be compared, after correction by a
1/$T$ factor due to the Curie behaviour of the nuclear magnetization
(thermal variations of the characteristics of the NMR circuit are much smaller than
the effects found here).
Inset: The two contributions of the room temperature $^{63}$Cu spectrum
at 17 and 24~Tesla.}
\label{CuNMR}
\end{figure}
To our knowledge, the only other situation which could generate an
inhomogeneous staggered
magnetization is the presence of magnetic clusters, such as would
be generated by finite size hole-free regions.
The corrolary of this is the presence of surrounding hole-rich regions.
Their exact topology cannot be inferred here, so we will call them "domain-walls".
In such a scenario, the main peak, which disappears at low $T$,
corresponds to hole-rich regions, {\it i.e.} where domain-walls are still mobile.
In fact, the wall-motion averages out $<$$S_z$$>$ (spin-flips), yielding a narrow
central peak. This also reduces the magnetic coupling between hole-poor domains.
The spatially inhomogeneous profile of $<$$S_z$$>$ within each domain
and the distribution of cluster sizes yield the broad background.
Full localization of domain-walls is likely to
restore inter-cluster magnetic coupling, thus enabling spin-freezing.
Of course, there must be significant disorder in the domain-wall
topology, in order to prevent long range AF ordering.
The disappearance of the main Cu peak is compatible 
with the localization of walls, which reduces the effective width
of hole-rich regions.
Accordingly, this peak disappears in the temperature region
where $\rho_{ab}$ becomes insulating-like.
The concomitant growth of $<$$S_z$$>$ explains the broadening
of the background signal.


\la~NMR spectra offer a second possibility to probe the phase
separation in CuO$_2$ planes.
A shown in Fig. 3, a second peak emerges upon cooling on the low frequency side
of the spectrum.
Qualitatively, we can ascribe the new peak to the
\la~nuclei within AF clusters, as a confirmation of the \cu~NMR spectra.
Similar experiments at 4.7~Tesla show a single peak (not shown),
with a $T$-dependent asymetry which is well-fitted by the sum of
two gaussians, whose separation is half of that at 9.4~T.
This again proves that the peaks are related to two different
{\it magnetic} environnements.
Additional magnetic broadening at low-$T$ makes the two \la~peaks unresolved,
and not surprinsingly, the broadening becomes noticable below $\sim$70~K,
where the spin fluctuations start to slow down.
Again, we stress that macroscopic doping inhomogeneities in the sample
would not produce such a $T$-dependence of the relative intensities of
the two NMR contributions. The observed phase separation clearly develops on
decreasing temperature.
\vspace{-1cm}
\begin{figure}[b!]
\begin{center}
 \epsfxsize=85mm
 $$\epsffile{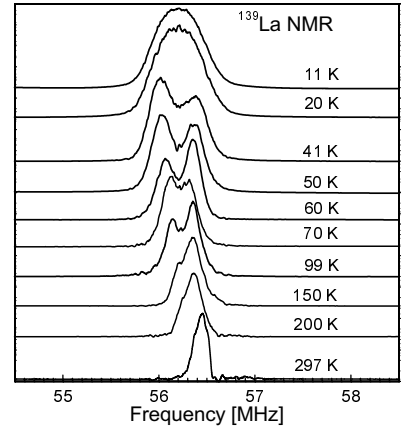}$$
\end{center}
\caption{$^{139}$La NMR spectra obtained by Fourier
transforms of the spin-echo ($H_0$=9.4~T, $\|$$c$).}
\label{LaNMR}
\end{figure}
Furthermore, similar \la~NMR results have been recently
obtained in La$_{1.9}$Sr$_{0.1}$CuO$_4$ \cite{Julien99b} and in
La$_2$CuO$_{4+\delta}$ at a concentration where long range
spin and charge ordering are absent \cite{Brom}.
This shows that the results are not unique to our Sr concentration.
Rather, phase separation appears to be a general
tendency in these materials.
In fact, most striking is probably the similarity between our \la~NMR spectra
and those reported in stripe-ordered nickelates \cite{Yoshinari98},
although details differ due to the difference of
hyperfine interactions, doping levels and
stripe configurations between cuprates and nickelates.


A quantitative analysis, like the comparison between \cu~and \la~
spectra, is however difficult since a number of Cu nuclei are not observed
and hyperfine interactions are not well known for \la~in the paramagnetic phase.
Furthermore, the relation of the \la~peak intensity ratio to the relative
size of the two phases is expected to be much more complex than the value
$\sim$1/16 determined by the hole concentration. Many microscopic details like the 
profile of the spin modulation and the organization (topology, filling)
of the hole-rich region are involved.
Even in the case of La$_{5/3}$Sr$_{1/3}$NiO$_4$, with established stripe order, 
the two-peak intensity ratio is not well-understood \cite{Yoshinari98}.


Fig. 4 summarizes our findings in \lsix: \cu~and \la~NMR spectra reveal that
magnetic phase separation develops below room temperature.
The data are best explained in terms of
hole-poor regions (AF clusters are evidenced through an anomalous NMR line)
and hole-rich regions (contributing a more usual line).
In the regime were doped-holes are localized ("charge glass"), the dynamics of
staggered moments, probed by NMR relaxation, slows down.
Below 5~K, {\it in the superconducting state}, AF clusters are frozen,
a phase called "cluster spin-glass".
Although no direct evidence for stripe-like objects is claimed here, the evidence
for their existence at somewhat higher doping ($x$$\simeq$0.12
\cite{Tranquada95,Tranquada97}) does suggest that
hole-rich regions are related to charge-stripes that are progressively pinned by
random (Sr) disorder as $T$ decreases. The charge-freezed state would then correspond
to a static disordered stripe phase: a "stripe-glass" \cite{Kivelson98,Boebinger96}.


\vspace{-1cm}
\begin{figure}[h!]
\begin{center}
 \epsfxsize=80mm
 $$\epsffile{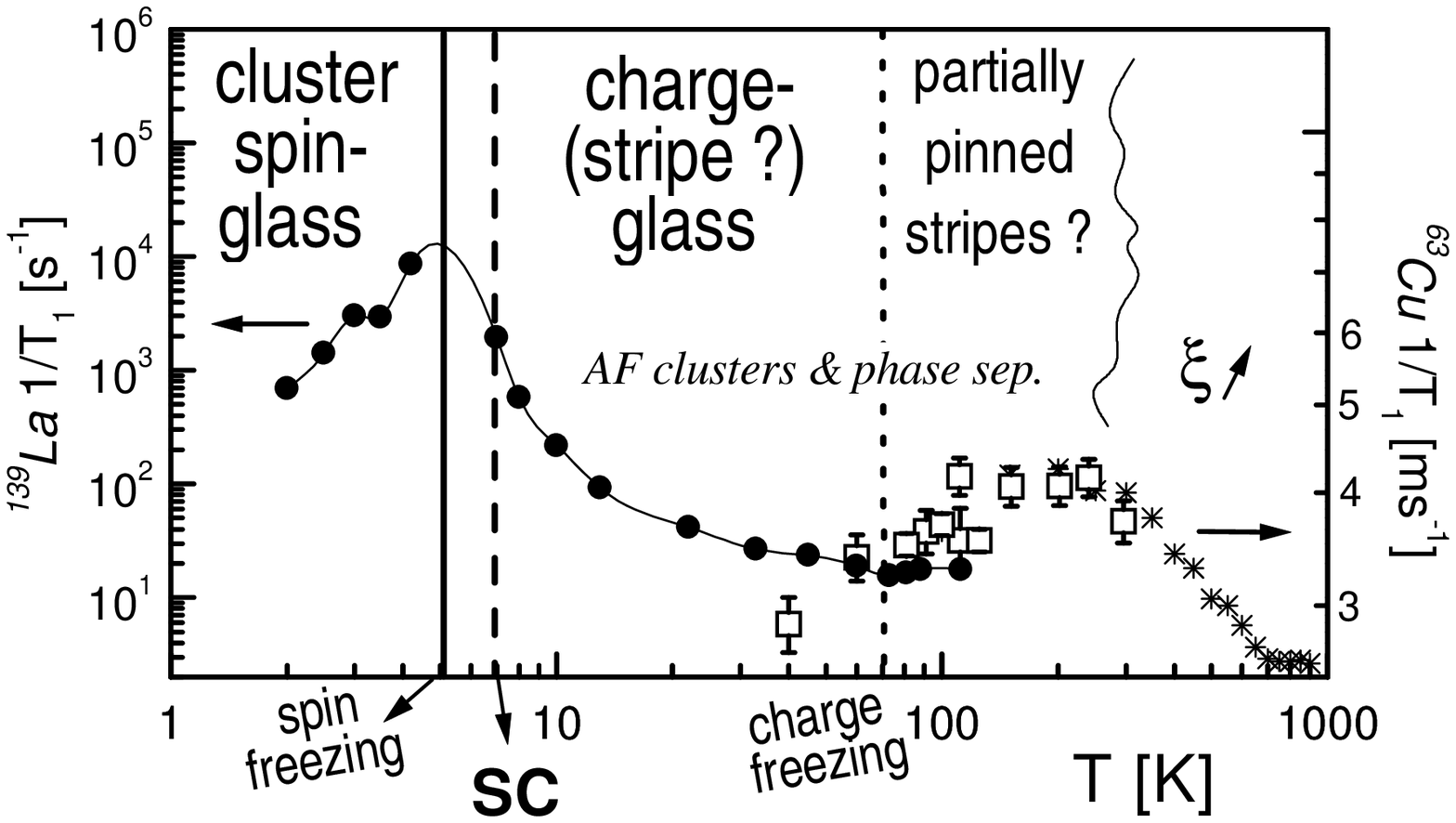}$$
\end{center}
\vspace{-7cm}
\caption{Experimental summary (crosses are data from
Ref.~\protect\cite{Fujiyama97}) and tentative phase diagram of \lsix.}
\label{summary}
\end{figure}
The above conclusions are further supported by:
1) the already mentioned similarities with NMR data in stripe-ordered materials,
2) the fact that even materials with well-established
stripe order tend to have a glassy behaviour \cite{Glass},
3) the presence of incommensurate elastic peaks in neutron scattering for
$x$=0.06 \cite{Wakimoto99},
4) the two-component ARPES spectra in the spin-glass region \cite{Ino99}.
This, to our knowledge first, observation of two-phases NMR
spectra in superconducting LSCO opens new perspectives:
Given the similarities between LSCO and YBCO \cite{Niedermayer98},
an NMR re-investigation of their underdoped regime is clearly called for.


Useful exchanges with H.B. Brom, V.J. Emery, R.J. Gooding,
P.C. Hammel, A. Rigamonti and B.J. Suh are acknowledged.
We thank S. Aldrovandi, Z.H. Jang, E. Lee, L. Linati
and F. Tedoldi for help, as well as J.E. Ostenson
and D.K. Finnemore for magnetization measurements.
The work in Pavia was supported by the INFM-PRA SPIS funding.
Ames Laboratory is operated for U.S Department 
of Energy by Iowa State University under Contract No. W-7405-Eng-82. The work at 
Ames Laboratory was supported by the director for Energy Research, Office of Basic 
Energy Sciences.


\vspace{-4mm}



\begin {references}
\bibitem{T1} The \la~$T_1$ values
reported here are defined as the time at which the nuclear magnetization has decreased
by a factor 1/$e$ from its equilibrium value; 
the value of $T_g$ is not influenced by the criterion chosen.
\end{references}
\end{document}